# Locating Restricted Facilities on Binary Maps

**Mugurel Ionut Andreica, Politehnica University of Bucharest, mugurel.andreica@cs.pub.ro**
**Cristina Teodora Andreica, Commercial Academy Satu Mare**
**Madalina Ecaterina Andreica, Academy of Economic Studies, Bucharest**

**Abstract:** *In this paper we consider several facility location problems with applications to cost and social welfare optimization, when the area map is encoded as a binary (0,1) $m \times n$ matrix. We present algorithmic solutions for all the problems. Some cases are too particular to be used in practical situations, but they are at least a starting point for more generic solutions.*

## 1 Introduction

Locating facilities which minimize costs or maximize social welfare have important applications in many situations. In this paper we consider the case in which the area map is encoded as a binary matrix A with m rows and n columns. The encoding could represent clean and polluted areas or desirable and undesirable zones. For this encoding, we consider several facility location problems to which we present optimal algorithmic solutions. Some of the problems are too particular or too restricted to be used in pratical situations directly, but we believe that they represent a starting point for handling more general problems. The rest of this paper is structured as follows. In Sections 2-5 we consider several facility location problems (largest squares, diamonds and rectangles). In Section 6 we compute depth arrangements of connected components in binary maps and in Section 7 we conclude. Related work is presented for each problem.

## 2 Locating Square Facilities with Varied Patterns

We want to find the largest square submatrix having a specific property. The case when we want to find a *monotone* square, whose cells are all equal (to 0 or 1) is part of folklore. We also consider two other patterns: the largest chessboard patterned square (where the values of the cells alternate on each row and column) and the largest square having all 1s on the main diagonal and 0s in the rest of the square. For the monotone square submatrix, we compute SQ(i,j)=the side of the largest square whose upper left corner is at row i and column j:

$$SQ(i,j) = \begin{cases} 1, & \text{if } (i=m) \text{ or } (j=n) \\ 1, & \text{if } (A(i+1,j) \neq A(i,j)) \text{ or } (A(i,j+1) \neq A(i,j)) \text{ or} \\ & (A(i+1,j+1) \neq A(i,j)) \\ 1+\min\{SQ(i+1,j), SQ(i,j+1), SQ(i+1,j+1)\}, & \text{otherwise} \end{cases} \quad (1)$$

For the largest „chessboard", we compute CB(i,j)=the largest chessboard with the upper right corner at (i,j):

$$CB(i,j) = \begin{cases} 1, & \text{if } (i=m) \text{ or } (j=n) \\ 1, & \text{if } (A(i+1,j) = A(i,j)) \text{ or } (A(i,j+1) = A(i,j)) \text{ or} \\ & (A(i+1,j+1) \neq A(i,j)) \\ 1+\min\{CB(i+1,j), CB(i,j+1), CB(i+1,j+1)\}, & \text{otherwise} \end{cases} \quad (2)$$

For the largest square with all 1s on the main diagonal and all 0s in the rest we compute IM(i,j)=the largest such square with the upper right corner at row i and column j:

$$IM(i, j) = \begin{cases} 0, \text{ if } A(i, j) = 0 \\ 1, \text{ if } ((i = m) \text{ or } (j = n)) \text{ and } (A(i, j) = 1) \\ 1, \text{ if } (A(i+1, j) = 1) \text{ or } (A(i, j+1) = 1) \text{ or } (A(i+1, j+1) = 0) \\ \min\{1 + IM(i+1, j+1), RM(i, j+1), DM(i+1, j)\}, \text{ otherwise} \end{cases} \quad (3)$$

RM(i,j) (DM(i,j)) is the length of the longest sequence of consecutive 0s starting at (i,j) and going right (down):

$$RM(i, j) = \begin{cases} 0, \text{ if } (A(i, j) = 1) \\ 1, \text{ if } (j = n) \text{ and } (A(i, j) = 0) \\ 1 + RM(i, j+1), \text{ if } (j < n) \text{ and } (A(i, j) = 0) \end{cases} \quad (4)$$

$$DM(i, j) = \begin{cases} 0, \text{ if } (A(i, j) = 1) \\ 1, \text{ if } (i = m) \text{ and } (A(i, j) = 0) \\ 1 + DM(i+1, j), \text{ if } (i < m) \text{ and } (A(i, j) = 0) \end{cases} \quad (5)$$

We can extend the patterns used for computing the largest squares specific properties to a more general case, where one can compute the following three values: RM(i,j)=$f_{RM}$(A(i,j), RM(i,j+1)), DM(i,j)=$f_{DM}$(A(i,j), DM(i+1,j)), BEST(i,j)=$f_{BEST}$(A(i,j), A(i+1,j), A(i,j+1), A(i+1,j+1), BEST(i+1,j), BEST(i,j+1), BEST(i+1, j+1)), where $f_{RM}$, $f_{DM}$ and $f_{BEST}$ are functions which can be evaluated in O(1) time. This way, we obtain O(m×n) algorithms for computing an optimal square submatrix consistent with the definitions of $f_{RM}$, $f_{DM}$ and $f_{BEST}$. In [1], all maximal monotone squares are efficiently found.

## 3 Locating Diamonds with Various Patterns

A diamond is similar to a square rotated by 45 degrees. A diamond with the center at row i and column j and with side length L consists of all the cells at positions

(p,q), such that |p-i|+|q-j|<L. We want to find the largest diamond (the one with the largest side length), which is composed of cells equal to 1. A good approach is to decompose a diamond into elementary shapes whose optimal sizes can be computed quickly. We notice that a diamond with its center at (i,j) can be considered as the (non-disjoint) union of four isosceles, right-angled triangles, with their right angle at (i,j) (the four triangles are oriented towards north-west, north-east, south-east and south-west). We now show how we can compute TSE(i,j)=the largest triangle full of 1s oriented towards south east, with its right angle at position (i,j). The triangles oriented towards other directions are computed similarly. We notice that a south-east oriented isosceles triangles full of 1s is equivalent to a square in which the cells on and above the secondary diagonal are 1s and we don't care about the other cells:

$$TSE(i, j) = \begin{cases} 0, \text{ if } A(i, j) = 0 \\ 1, \text{ if } (A(i, j) = 1) \text{ and } ((i = m) \text{ or } (j = n) \text{ or } \\ \quad (A(i+1, j) = 0) \text{ or } (A(i, j+1) = 0) \text{ or } (A(i+1, j+1) = 0)) \\ 1 + \min\{TSE(i, j+1), TSE(i+1, j), TSE(i+1, j+1)\}, \text{ otherwise} \end{cases} \quad (6)$$

The side of the largest diamond with its center at cell (i,j) is min{TSE(i,j), TSW(i,j), TNE(i,j), TNW(i,j)}, where TSW(i,j), TNE(i,j) and TNW(i,j) are the sides of the largest isosceles right-angled triangles with their right angle at (i,j) and oriented towards south-west, north-east or north-west. A simpler method consists of computing D(i,j)=the side of the largest diamond with the uppermost cell (the one on the smallest numbered row) at (i,j):

$$D(i, j) = \begin{cases} 0, \text{ if } A(i, j) = 0 \\ 1, \text{ if } (A(i, j) = 1) \text{ and } ((i = m) \text{ or } (j = n) \text{ or } \\ \quad (A(i+1, j-1) = 0) \text{ or } (A(i+1, j) = 0) \text{ or } (A(i+1, j+1) = 0)) \\ 1 + \min\{D(i+1, j-1), D(i+1, j), D(i+1, j+1)\}, \text{ otherwise} \end{cases} \quad (7)$$

The O($m_x$n) algorithm can be generalized by defining a function $f_D$ such that D(i,j)=$f_D$(A(i,j), A(i+1,j-1), A(i+1,j), A(i+1,j+1), D(i+1,j-1), D(i+1,j), D(i+1,j+1)).

## 4 Largest Area Empty Rectangle

We are given n points in the plane, included inside an axis-parallel rectangle D=[0,$X_{max}$]x[0,$Y_{max}$]. We want to find an axis-parallel rectangle with the largest possible area, which is included inside the rectangle D and contains no point inside it (only, eventually, on its contour). This problem has been solved in O(n·$\log^2$(n)) [2], but we propose here a very simple O($n^2$) algorithm. We will construct a binary matrix, in the following way. We sort the x-coordinates of the points: $x_1 \leq x_2 \leq ... \leq x_n$. We add $x_0$=0 and $x_{n+1}$=$X_{max}$. We then do the same with the y-coordinates (with $y_{n+1}$=$Y_{max}$). We (logically) construct a matrix A with 2·n+1 rows and 2·n+1 columns where each row i (0≤i≤2·n) has a possibly different height $rh_i$ and each column j (0≤j≤2·n) has a possibly different width $cw_j$: $cw_0$=$x_1$-$x_0$, $cw_{2·n}$=$x_{n+1}$-$x_n$ and for 1≤j≤2·n-1: $cw_j$=0, if j is odd, or $x_{(j/2)+1}$-$x_{(j/2)}$, if j is even; we define the row heights similarly. If a point has x- and y-coordinates ($x_j$, $y_i$), 1≤i,j≤n, the matrix will contain a 1 at the following (row, column): (2·i-1, 2·j-1); all the other cells will contain 0s. We will find the maximum area rectangle full of 0s in this binary matrix. We consider each row i and try to find the largest area rectangle with its lower side on row i. For each column j on row i, we compute its height $ch_{i,j}$=the sum of row heights of all the consecutive cells on column j, up to the first cell containing a 1 or the matrix border (if A(i,j)=1 then

$ch_{i,j}=0$, else $ch_{i,j}=rh_i+ch_{i-1,j}$, where $ch_{-1,j}=0$). With these values, we use the staircase technique [3] for each cell (i,j) in the matrix. The difference consists of the fact that the width of a column j is $cw_j$ instead of 1.

## 5 Maximum Perimeter Sum Rectangle

In this section we consider a matrix B with real values (positive or negative) and we want to find a subrectangle with the maximum perimeter sum (sum of the values of the cells on the perimeter). The subrectangle must have at least 2 rows and 2 columns. We will consider every pair of rows $l_1$ and $l_2$ ($l_1<l_2$) and determine the optimal rectangle with its upper side on row $l_1$ and its lower side on row $l_2$. By doing this, we transform the problem into a one-dimensional one. We define an array $v(l_1,l_2)$ with n cells, where $v(l_1,l_2)(j)$ is the sum of the values on column j, considering only the rows between $l_1$ and $l_2$ (inclusive). In order to compute these values fast, we will pre-compute a matrix of partial sums: $S_{col}(i,j)$=the sum of the values on column j considering the rows 1, 2, ..., i. We have $S_{col}(0,j)=0$ and $S_{col}(i\geq1,j)=B(i,j)+S_{col}(i-1,j)$. With this partial sums matrix, we have $v(l_1,l_2)(j)=S_{col}(l_2,j)-S_{col}(l_1-1,j)$. We will compute $w(l_1,l_2)(j)$=the largest perimeter sum of a rectangle with its right side on column j and $u(l_1,l_2)(j)$=the largest perimeter sum of a „quasi-rectangle" (with its left side at a column k≤j and extending all the way to column j, but column j is not its right side). We have $u(l_1,l_2)(1)=v(l_1,l_2)(1)$ and $u(l_1,l_2)(j>1)=\max\{v(l_1,l_2)(j),\ u(l_1,l_2)(j-1)+B(l_1,j)+B(l_2,j)\}$; $w(l_1,l_2)(j)=u(l_1,l_2)(j-1)+v(l_1,l_2)(j)$. The maximum value among all $w(l_1,l_2)(j)$ ($1\leq l_1<l_2\leq m$, $1\leq j\leq n$) is the maximum perimeter sum of a rectangle. The maximum sum subarray problem [4] is related to this problem.

## 6 Depth Arrangements of Connected Components

A black connected component in a binary matrix is a maximal set of cells all equal to 1, such that we can travel from any one of them to any other using only the horizontal and vertical directions (simple set of movements) and passing only through cells belonging to the component. A white connected component is defined similarly, but we can use the horizontal, vertical and diagonal directions (extended set of components). The black components may enclose some other black components; thus, we define the depth of a black component as 1 plus the number of black components enclosing it. We will compute a matrix $D(i,j)$=the depth of the cell located on row i and column j. We initialize all the white cells on the border of the matrix with depth 1 and insert them in a double-ended queue Q (if all the border cells are black, we initialize their depth with 2 and insert them into Q); the depth of the other cells is left unitinitialized and now we use the following algorithm:

**DepthComputation():**
*initialize Q with the border cells (white or black)*
**while** (*Q≠empty*) **do**
 *(r,c)=Q.removeFromTheFront()*
 **for** *(r',c') neighbor of (r,c) such that D(r',c')=uninitialized* **do**
  **if** (*A(r',c')=A(r,c)*) **then**
   *D(r',c')=D(r,c); Q.insertAtTheFront((r',c'))*
  **else**
   *D(r',c')=D(r,c)+1; Q.insertAtTheEnd((r',c'))*

The neighbors of a black cell are defined using the simple set of movements and those of a white cell using the extended set. With this algorithm, all the cells of a black (white) component have the same depth and the depth of a black component is $D(i,j)/2$ (where $(i,j)$ is a cell of the component). We can also detect black components which enclose at least one white cell, if there exists a black cell $(i,j)$ of the component and a neighboring white cell $(i',j')$ such that $D(i',j')=D(i,j)+1$. We can use the same algorithm if we define black connectivity using the extended set of movements and white connectivity using the simple set.

## 7 Conclusions and Future Work

In this paper we presented several (generic) algorithmic solutions to some facility location problems on maps encoded as binary matrices. The problems we considered have applications to cost and social welfare optimization, although some of them cannot be used in practical situations immediately, due to the particular nature of the problem being solved. As future work, we intend to tackle more difficult problems, with more direct applications to real life requirements.